\documentclass[prb,twocolumn,showpacs,showkeys,superscriptaddress]{revtex4}%
\usepackage{times}
\usepackage{graphicx}
\usepackage{dcolumn}
\usepackage{bm}
\usepackage[usenames,dvipsnames]{color}
\usepackage{amsmath}
\usepackage{amsfonts}
\usepackage{amsthm}
\usepackage{amssymb}
\usepackage{mathrsfs}
\usepackage{subfigure}
\usepackage{ulem}
\usepackage{verbatim}

\begin{document}

\title{Open quantum system description of singlet-triplet qubits in quantum dots}

\author{L. K. Castelano}
\email{lkcastelano@ufscar.br}
\affiliation{Departamento de F\'{\i}sica, Universidade Federal de S\~ao Carlos, 13565-905, S\~ao Carlos, SP, Brazil}
\author{F. F. Fanchini}
\affiliation{Faculdade de Ci\^encias, UNESP - Universidade Estadual Paulista, Bauru, SP, 17033-360, Brazil}
\affiliation{The Abdus Salam International Centre for Theoretical Physics, Strada Costiera 11, Miramare-Trieste, Italy}
\author{K. Berrada}
\affiliation{The Abdus Salam International Centre for Theoretical Physics, Strada Costiera 11, Miramare-Trieste, Italy}

\begin{abstract}
We develop a theoretical model to describe the dissipative dynamics of singlet-triplet ($S$-$T_0$) qubits in GaAs quantum dots. Using the concurrence experimentally obtained as a guide, we show that each logical qubit is coupled to its own environment because the decoherence effect can be described by independent dephasing channels. Given the correct description of the environment, we study the dynamics of concurrence as a function of the temperature, the constant coupling between the system and the environment, the preparation time, and the exchange coupling. Although the reduction of the environment coupling constant modifies the entanglement dynamics, we demonstrate that temperature emerges as a crucial variable and a variation of  millikelvins significantly modifies the generation of entangled states. Furthermore, we show that the exchange coupling together with the preparation time strongly affects the entanglement dissipative dynamics.

\end{abstract}

\date{\today}
 \maketitle

\section{Introduction}
The development of quantum information processing have discovered different new techniques which are paving the way to accomplish quantum technological advancement.~ \cite{newt}
Among these advances, spin qubits in quantum dots (QDs) is certainly one of the most striking systems because of their potential scalability and miniaturization.~\cite{scale1,scale2,scale3} Furthermore, electrical readout and control of single spins in quantum dots (QDs) have been proven very challenging, 
where spin blockade and charge sensors enable the observation of single/two-spin dynamics.\cite{qd1,qd2}

More recently, a significant progress in implementing an alternative
scheme in double quantum dots (DQDs) has been attained.~ \cite{qdn1} In this approach, singlet and triplet states of two electrons are used to represent a logical qubit and, by means of this apparatus, a breakthrough experiment shows that a universal set of quantum gates can be reached.~\cite{sci} Furthermore, the inter-qubit interaction can be implemented through a capacitively mediated dipole-dipole coupling that exploits differences between charge configurations of singlet and triplet states to control the interaction between logical qubits.

In this work, we present a theoretical model to describe the dynamics of two singlet-triplet qubits interacting with the environment. By using such a model, we are able to quantitatively reproduce experimental results; thereby, understanding how singlet-triplet qubits in GaAs quantum dots interact with their environment. Moreover, we show that qubits are weakly coupled to independent dephasing channels and we reproduce the entanglement dissipative dynamics observed experimentally in Ref.~[\onlinecite{sci}]. Such results can be used as a reference for further studies of open quantum system based on QDs. We study the entanglement dynamics as a function of the temperature, which plays an important role in the characterization of entanglement. Finally, with the correct description of the open quantum system, we exploit the role of the preparation time, \textit{i.e.} the time to prepare the necessary initial superposed state, in the entanglement dissipative dynamics.

The present paper is organized as follows: In Sec. II we present the model that describes the dynamics of the open quantum system based on DQDs. In Sec. III, we present the time local second-order master equation used to determine the reduced dynamics of the quantum states. In Sec. IV, we present the concept of entanglement, measured by concurrence together with results of our theoretical model including a detailed study concerning the coupling constant, the temperature, the environment structure, and the role of the preparation time in the entanglement dynamics. Finally, Sec. V contains a summary of our results.

\section{Theoretical Model}
The main focus of this work is related to the study of the dissipative dynamics of a pair of singlet-triplet ($S$-$T_0$) qubits, where the information is stored in the spin state of two electrons. Such states can be experimentally achieved by confining two electrons in each DQD system.~\cite{sci} Moreover, the two-level system of a logical qubit ($|S\rangle\equiv|\uparrow\rangle$,$|T_0\rangle\equiv|\downarrow\rangle$) can be isolated by applying an external magnetic field in the plane of the device in such a way that the Zeeman splitting makes the parallel spin states $|T_+\rangle$ and $|T_-\rangle $ energetically inaccessible.

To extend such a two-level system to a two-qubit system, it is necessary to couple two $S$-$T_0$ qubits, where the tunnelling between them is suppressed and their coupling is electrostatic (for more details, see ref.~[\onlinecite{sci}]). Thus, the effective Hamiltonian for two-qubits system can be written as follows:~\cite{sci}     
\begin{eqnarray}
\hat{H}_{\text{2-qubit}}&=&{1\over2}\Bigg[\left(J_1\sigma_z^{(1)}\otimes \mathbf{I}+J_2\;\mathbf{I}\otimes \sigma_z^{(2)}\right)+\\\nonumber&&{J_{12}\over2}\left(\sigma_z^{(1)}\otimes
\sigma_z^{(2)}-\sigma_z^{(1)}\otimes \mathbf{I}-\mathbf{I}\otimes\sigma_z^{(2)}\right)+\\\nonumber&&{1\over2}\left(\Delta B_{z,1}\sigma_x^{(1)}\otimes\mathbf{I} +\Delta B_{z,2}\;\mathbf{I}\otimes \sigma_x^{(2)}\right)\Bigg],
\end{eqnarray}
where $\sigma_{x,y,z}$ are the Pauli spin matrices, $\mathbf{I}$ is the identity and the index 1 (2) is related to first (second) qubit (hereafter, we use units of $\hbar =1$). This Hamiltonian is able to implement universal quantum control, which is given by two physically distinct local operations, $x$ and $z$, and by the interaction between the qubits given by $\sigma_z^{(1)}\otimes\sigma_z^{(2)}$. The exchange splitting, $J_{i}$, between $|S^i\rangle$ and $|T^i_0\rangle$ applies rotations in the qubit $i$=1,2 around the $z$ axis, while rotations around $x$ axis are driven by a magnetic field gradient $\Delta B_z$.  Moreover, $\Delta B_z$ is responsible to prepare each qubit in a superposition between $|S\rangle$ and $|T_0\rangle$. The exchange splitting, $J_{12}$, depends on the energy between levels of the left and the right DQD and it can be switched on and off during the quantum dynamics.~\cite{sci}  When $J_{12}$ is different from zero, because of the Pauli exclusion principle, the $|S\rangle$ and $|T_0\rangle$ states have different charge configuration, which makes the state of the first qubit to be conditioned to the state of the second qubit. In other words, when simultaneously evolving, they experience a dipole-dipole coupling that generates an entangled state. Following the experimental steps, \cite{sci} each qubit is initialized in the $|S\rangle$-state, then  rotated by $\pi/2$ around the $x$ axis when $J_i=J_{12}=0$, $\Delta B_{z,i}/2\pi\approx 30 \rm {MHz}$, for i=1,2. After this stage, a large exchange splitting is switched on corresponding to $J_1/2\pi\approx 280 {\rm {MHz}}$, and $J_2/2\pi\approx 320 {\rm {MHz}}$. Experimentally, it was found that the two-qubit coupling is given by $J_{12}= J_1J_2$.~\cite{sci}

To include the dissipative dynamics, we suppose
that qubits are coupled to a bath of harmonic oscillators.~\cite{file} Such a coupling induces dephasing channels, which are the main source of dissipation in quantum dots.~\cite{loss}  Thus, the total
Hamiltonian that computes the environment contribution is given by
\begin{equation}
\hat{H}=\hat{H}_{\text{2-qubit}}+\hat{H}_b+{\hat{H}}_{{int}}
\end{equation}
where $\hat H_b$ is the bath Hamiltonian and $H_{int}$ is the interaction Hamiltonian. In this work, we want to determine if the environment interacts collectively or independently with the system, \textit{i.e.}, if both qubits noise are correlated or not. To verify such correlations, we suppose two distinct interaction Hamiltonian. First, the interaction Hamiltonian is considered as a common dephasing environment 
\begin{equation}
\hat{H}_{\text{int}}^C=\left(\sigma_z^{(1)}+\sigma_z^{(2)}\right)\mathcal{L},\label{Hcol}
\end{equation}
where $\mathcal{L}=B+B^\dag$ with $B=\sum_kg_kb_k$ and $g_k$ is a complex coupling constant. In the common environment case, both qubits interact with a common bath,  whose Hamiltonian is given by
 \begin{equation}
 \hat{H}_b^C=\sum_k\omega_kb_k^\dag b_k,\label{Hbcol}
 \end{equation}
 where $\omega_k$ is the frequency of the ${k}^{\text{th}}$ normal mode of the bath and $b^\dag_k$ ($b_k$)
is the creation (annihilation) operator of the reservoir field. Superscripts of $ \hat{H}_b^C$ and $\hat{H}_{\text{int}}^C$ refer to collective bath.

 In the second case, the interaction Hamiltonian is composed of independent dephasing environments for each qubit as follows 
\begin{equation}
\hat{H}_{\text{int}}^I=\sigma_z^{(1)}\mathcal{L}^{(1)}+\sigma_z^{(2)}\mathcal{L}^{(2)}\label{Hind},
\end{equation}
where $\mathcal{L}^{(i)}=B^{(i)}+{B^\dag}^{(i)}$ with $B^{(i)}=\sum_kg_k^{(i)}b_k^{(i)}$ and $g_k^{(i)}$ is a complex coupling constant of the $i^\text{th}$ qubit. In this case, we have two independent baths described by
 \begin{equation}
 \hat{H}^{I}_b=\sum_{i=1}^2\sum_k\omega_k^{(i)}{b_k^\dag}^{(i)} b_k^{(i)},\label{Hbind}
 \end{equation}
 where $\omega_k^{(i)}$ is the frequency of the ${k}^{\text{th}}$ normal mode of the $i^\text{th}$ bath and ${b^\dag_k}^{(i)}$ ($b_k^{(i)}$) is the creation (annihilation) operator of the $i^\text{th}$ reservoir field. Superscripts of $ \hat{H}_b^I$ and $\hat{H}_{\text{int}}^I$ refer to independent baths.
 The purpose of considering independent and common environment is to find a better physical model for the dynamics of the two DQDs open quantum system. By comparing our theoretical results with experimental results, we are able to extract such informations.
 
\section{Master Equation} 
To calculate the dissipative dynamics, we consider a time-local second-order master equation, which is given by
\begin{eqnarray}
\frac{d\rho _{I}(t)}{dt}=-\int^{t}_{0}dt^{\prime} {\rm Tr}_{B}\left\{{\left[H_{I}(t),\left[H_{I}(t^{\prime}),\rho _{B}\rho _{I}(t)\right]\right]}\right\},\label{master}
\end{eqnarray}
where $\rho _{I}(t)$ is the reduced density matrix for two DQDs, $H_{I}(t)$ is the interaction Hamiltonian in the interaction picture, namely,
$H_{I}(t)=U^{\dagger}(t)U^{\dagger}_{B}(t)\hat H_{\rm int}U_{B}(t)U(t)$, with
$U_{B}(t)=\exp\left(-i\hat H_{b}t\right)$ and $U(t)=\exp\left(-i\hat H_\text{2-qubit}t\right)$. Equation (\ref{master}) is valid in the regime in which the strength of the coupling, expressed in frequency units, multiplied by the correlation time of the bath operators is much less than unity. To consider a collective environment in Eq.(\ref{master}), we must employ the interaction Hamiltonian and the bath Hamiltonian given by  Eqs.~(\ref{Hcol}) and (\ref{Hbcol}); on the other hand, Eqs.~ (\ref{Hind}) and (\ref{Hbind}) must be used for independent environments. We also suppose that the oscillator bath density matrix $\rho_B$ is initially decoupled from the system and it is given by 
\begin{eqnarray}
\rho _{B}=\frac{1}{Z}\exp(-\beta \hat H_{b}),\label{rhoB}
\end{eqnarray}
where $Z={\rm Tr}_{B}\left[\exp(-\beta \hat H_{b})\right]$ is the partition function. Here, $\beta =1/(k_{B}T)$, $k_{B}$ is Boltzmann constant, and $T$ is the absolute temperature of the environment. The bath of
oscillators is characterized by its spectral density 
that, in the limit where the number of bath normal modes per unit frequency becomes infinite, it can be defined as \cite{caldeira} 
\begin{eqnarray}
J(\omega )=\eta \omega \exp(-\omega / \omega _{c})\label{spectral},
\end{eqnarray}
where $\eta$ is a dimensionless constant coupling that describes the strength of the interaction between system and environment and $\omega _{c}$ is a cutoff frequency. 

\section{Results}
In this work, we use the quantum correlation called concurrence to perform the analysis of our results. Concurrence is a well known measure of entanglement, which is broadly accepted to be responsible for a set of important tasks in quantum information theory, such as quantum teleportation \cite{tele} and quantum key distribution \cite{key}. For two qubits, there is an analytical solution to concurrence,\cite{concurrence}which is given by
\begin{equation}
C(\rho)=\max\{0,\sqrt{\lambda_1}-\sqrt{\lambda_2}-\sqrt{\lambda_3}-\sqrt{\lambda_4}\},
\end{equation}
where $\lambda_i$ are the eigenvalues of $\rho\tilde{\rho}$ listed in descending order. $\tilde{\rho}$
is the time-reversed density operator,
\begin{equation}
\tilde{\rho}=(\sigma^{(1)}_y\otimes\sigma^{(2)}_y)\rho^*(\sigma^{(1)}_y\otimes\sigma^{(2)}_y),
\end{equation}
where $\rho^*$ is the conjugate of $\rho$ in the standard basis of two qubits.

The initial state of each DQD is set to $|\uparrow\rangle$, then a $\pi/2$ rotation around $x$-axis is performed during the preparation time $t_{pre}$, which put both qubits in a superposed state $|\uparrow\rangle+|\downarrow\rangle$. Following the experimental description given in ref.~[\onlinecite{sci}], we use $T=50$ mK, $\Delta B_{z,1}=\Delta B_{z,2}= (\pi/16\times 10^3) \rm {MHz} $, $J_1/2\pi = 280 {\rm {MHz}}$, and $J_2/2\pi = 320 {\rm {MHz}}$. The system dynamics can be obtained by numerically solving the master equation (Eq.~(7)). We perform a systematic analysis to characterize the environment, \textit{i.e.}, we seek values for the cutoff frequency $\omega_c$, for the constant coupling $\eta$, and for the common or the independent character of the environment that better describes the experimental data presented in ref.~[\onlinecite{sci}]. To determine these characteristics, we assume a weak coupling between both DQDs and the environment. In such a regime, the dissipative process is Markovian and the cutoff frequency is higher than other controllable field frequencies. After detailed analysis (not shown here) we find, as expected, that low cutoff frequencies are unable to reproduce the experimental data and we checked that all results presented in this work do not change significantly if $\omega_c> 2\times 10^4 \rm{MHz}$. Thus, the cutoff frequency $\omega_c=  2\times 10^4 \rm{MHz}$ is fixed hereafter.

\begin{figure} 
\includegraphics[width=8cm]{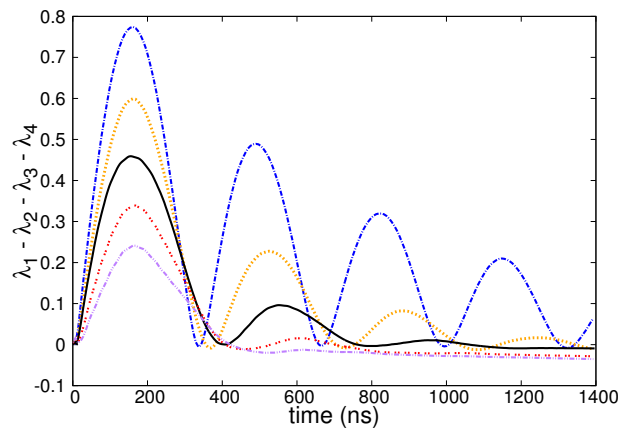}
\caption{\label{fig1} (Color online) Numerical solutions for the DDSE of the matrix $\rho\tilde\rho$ as a function of time, for a fixed temperature $T=50$ mK and a fixed cutoff $\omega_c=2\times 10^4 \rm{MHz}$, considering a common environment coupled to both qubits. The dot-dashed (blue) line represents the DDSE for $\eta=10^{-5}$, while the dotted (orange), the solid (black), the double dotted (red), and the double dot-dashed (magenta) respectively correspond to $\eta=2\times 10^{-5}$, $\eta=3\times 10^{-5}$, $\eta=4\times 10^{-5}$, and $\eta=5\times 10^{-5}$.}
\end{figure}

\begin{figure} 
\includegraphics[width=8cm]{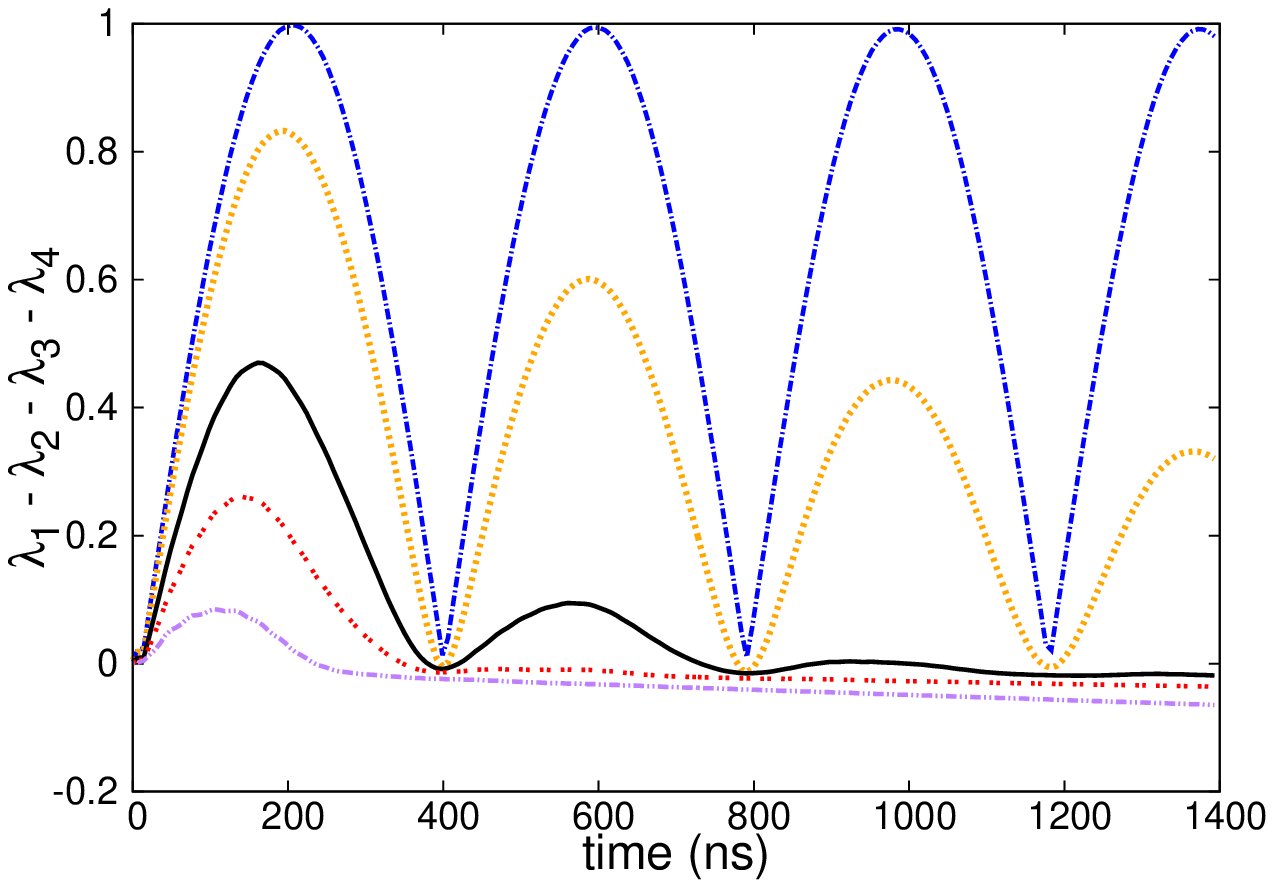}
\caption{\label{fig2} (Color online) Numerical solutions for the DDSE of the matrix $\rho\tilde\rho$ as a function of time, for a constant coupling $\eta=3\times 10^{-5}$ and a fixed cutoff $\omega_c=2\times 10^4 \rm{MHz}$, considering a common environment coupled to both qubits. The dot-dashed (blue) line represents the DDSE for $T=0$ mK, while the dotted (orange), the solid (black), the double dotted (red), and the double dot-dashed (magenta) respectively correspond to $T=10$ mK, $T=50$ mK, $T=100$ mK, and $T=200$ mK.}
\end{figure}

\begin{figure}
\includegraphics[width=8cm]{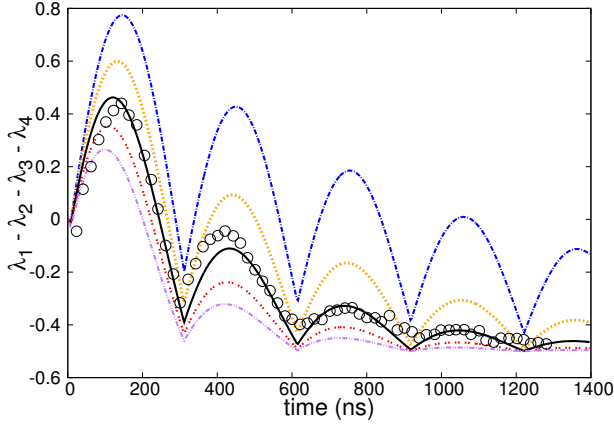}
\caption{\label{fig3} (Color online) Numerical solutions for the DDSE of the matrix $\rho\tilde\rho$ considering independent environments for each qubit, as a function of time for a fixed temperature $T=50$ mK and a fixed cutoff $\omega_c=2\times 10^4 \rm{MHz}$. The dot-dashed (blue) line represents the DDSE for $\eta=10^{-5}$, while the dotted (orange), the solid (black), the double dotted (red), and the double dot-dashed (magenta) respectively represent the results for $\eta=2\times 10^{-5}$, $\eta=3\times 10^{-5}$, $\eta=4\times 10^{-5}$,  and $\eta=5\times 10^{-5}$. Open circles denote the DDSE extracted from experimental data.~\cite{sci}}\label{both}
\end{figure}

\begin{figure} 
\includegraphics[width=8cm]{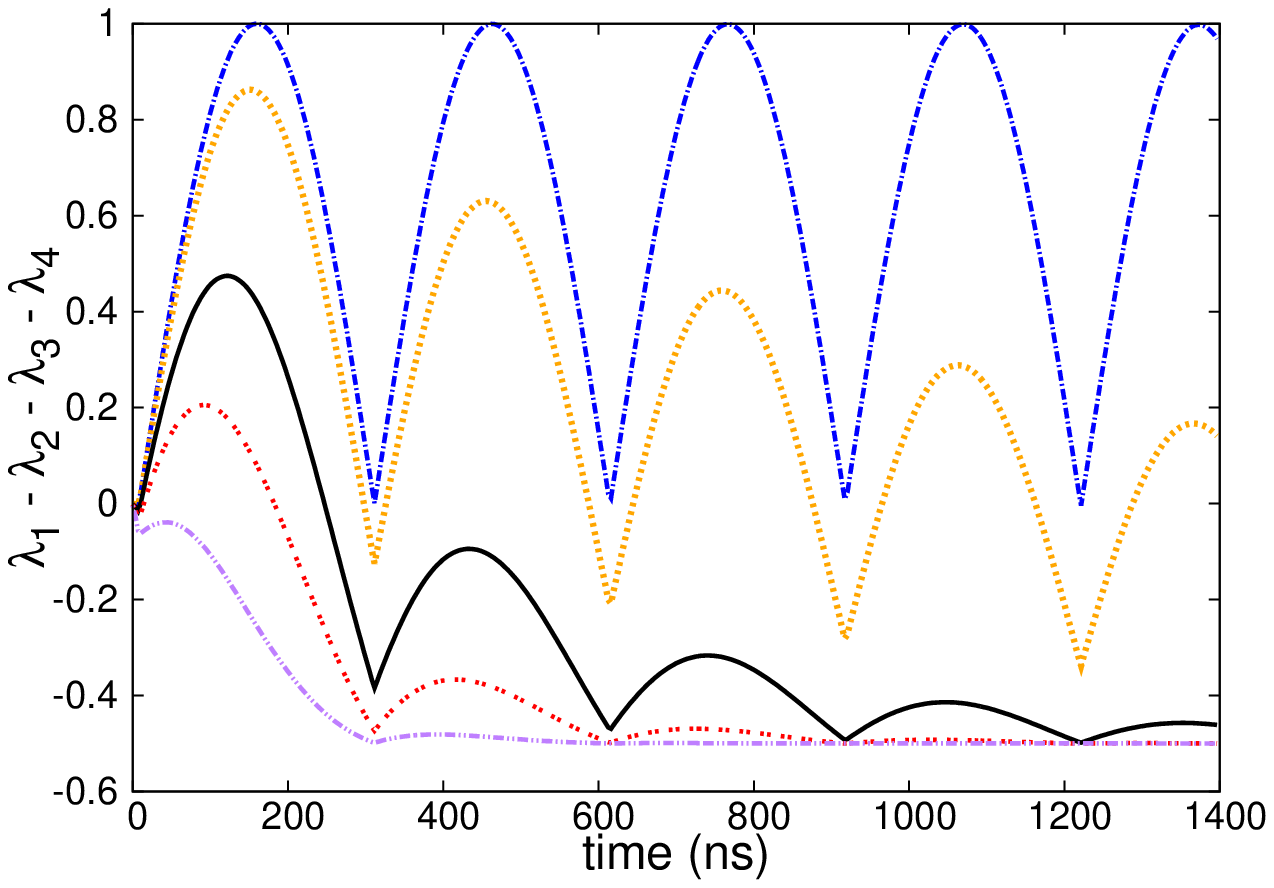}
\caption{\label{fig5}(Color online) Numerical solutions for the DDSE of the matrix $\rho\tilde\rho$ as a function of time, for a constant coupling $\eta=3\times 10^{-5}$ and a fixed cutoff $\omega_c=2\times 10^4 \rm{MHz}$, considering independent environments for each qubit. The dot-dashed (blue) line represents the DDSE for $T=0$ mK, while the dotted (orange), the solid (black), the double dotted (red), and the double dot-dashed (magenta) respectively represent the results for $T=10$ mK, $T=50$ mK, $T=100$ mK, and $T=200$ mK.}
\end{figure}

\begin{figure} 
\includegraphics[width=8cm]{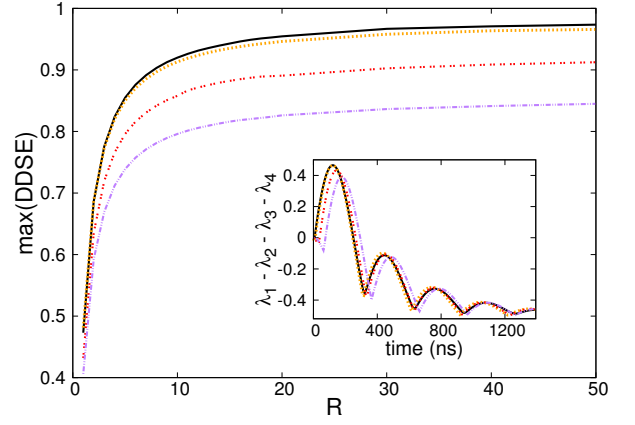}
\caption{\label{fig4} Numerical solutions for the maximum value of DDSE, considering independent environments, as a function of the amplitude of $R=J_{12}/J^{exp}_{12}$ and for different preparation times $t_{prep}$. The solid (black) is the result for $t_{prep}= 4$ ms, while the dotted (orange), the double-dotted (red), and the double dotted-traced respectively show the results for $t_{prep} = 8$ ms, $t_{prep} = 32$ ms, and $t_{prep} = 64$ ms. In the inset, we plot the DDSE of the matrix $\rho\tilde\rho$ as a function of time following the same preparation times and curve description used in the main panel.}\label{plotc}
\end{figure}

We begin our analysis by comparing the dynamics for the difference of the descending sorted eigenvalues (DDSE) $\lambda_1-\lambda_2-\lambda_3-\lambda_4$ of the matrix $\rho\tilde\rho$, which is equal to the concurrence $C(\rho)$ when it assumes positive values. 
In Fig.~(\ref{fig1}), we plot $\lambda_1-\lambda_2-\lambda_3-\lambda_4$ as function of time, considering a common environment coupled to both qubits and assuming different values for the coupling constant $\eta$, for the experimental temperature $T=50$ mK. One can see in Fig.~(\ref{fig1}) that the DDSE oscillates on time and it has an envelope function which decays faster as the constant coupling is increased. We also perform an analysis of DDSE as a function of time for a fixed coupling constant $\eta=3 \times 10^{-5}$ and different temperatures, which is shown in Fig.~(\ref{fig2}). The increasing of the temperature has a similar effect as the increasing of the constant coupling (Fig.~(\ref{fig1}); \textit{i.e.}, the higher the temperature, the faster the decay of the envelope function as a function of time.
 Based on results of Fig.~(\ref{fig1}) and Fig.~(\ref{fig2}), we conclude that the description of common environment for both DQDs does not describe the experimental results observed in ref.~[\onlinecite{sci}] because the DDSE do not present negative values, as found experimentally. Furthermore, the concurrence (positive values of DDSE) theoretically obtained presents a sudden-birth \cite{bellomo} which is not observed in the experiment. 

To provide a better description of such a experimental result, we also assume independent environments for each qubit. We choose the experimental temperature $T=50$ mK, in Fig.~(\ref{fig3}),  and we plot the DDSE as function of time, considering different coupling constants. Remarkably, for $\eta=3 \times 10^{-5}$, $\omega_c>2\times 10^4 \rm{MHz}$, there is a very good matching between experimental results (open circles in Fig.~(\ref{fig3})) and DDSE extracted from the reduced dynamics obtained for independent environments for each qubit. The experimental data for the DDSE, plotted in Fig.~(\ref{fig3}), achieves its largest value DDSE$\approx 0.45$ for $t=150$ ns; also, concurrence (positive values of the DDSE) is null for $t>250$ ns. The time where the concurrence is maximum can be related to term $J_{12}\sigma_z^{(1)}\otimes\sigma_z^{(2)}$ that couples both qubits, thus its amplitude $J_{12}$ is connected to the time where the maximum entanglement (concurrence) occurs. Such relation is $J_{12} = \pi/\tau_{ent}$, which gives $\tau_{ent}=150 {\rm {ns}}$ \cite{sci} in such a case.

\begin{figure} 
\includegraphics[width=8cm]{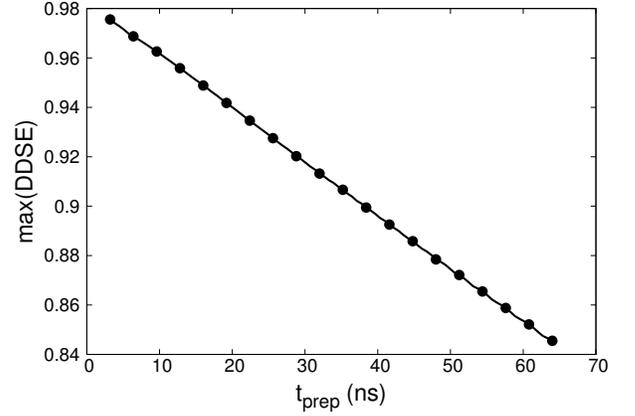}
\caption{\label{fig6}  Numerical solutions for the maximum value of DDSE as a function of $t_{prep}$ for $T=50$ mK, $\eta=3\times 10^{-5}$, $\omega_c=2\times 10^4 \rm{MHz}$, $R = 50$, and independent environments assumption.}
\end{figure}

By means of the correct description of the environment, we can analyze the entanglement dissipative dynamics through our theoretical model. We begin our analysis focusing on the role of the temperature in the dissipative process. In  Fig.~(\ref{fig5}), we fixed the parameters as the ones that better fit the experimental results and plot the DDSE as a function of time, for different temperatures. For $T=0$ K, we see that the electromagnetic vacuum has a negligible effect on the deterioration of the entanglement, which only oscillates as a function of time. Moreover, from Fig.~(\ref{fig5}), we see that a change of just $10$ mK significantly influences the entanglement dynamics and for $T=200$ mK, the entanglement definitely disappears.

We also study the role of the preparation time $t_{prep}$, \textit{i.e.} the time to prepare the superposed state. In the experiment, each qubit is initially prepared in the $|\uparrow\rangle$-state, then rotated by $\pi/2$ around the $x$-axis when $J_i=J_{12}=0$ to achieve the superposed state$|\uparrow\rangle+|\downarrow\rangle$. This rotation around $x$-axis is driven by a magnetic field gradient $\Delta B_z$. During the preparation time, the system is interacting with the environment, thereby affecting the entanglement dynamics. To probe such an effect, we study the role of $t_{prep}$ in the maximum value of entanglement during the dissipative dynamics. In the experimental result, for example, we have $t_{prep}\approx 8$ ns, $J^{exp}_{12}\approx\pi/150$ns, and the maximum obtained entanglement is around $0.45$ for $T=50$ mK. Naturally, another crucial aspect to maximize the entanglement between qubits is the exchange coupling between each DQD. The exchange coupling $J_{12}$, which can be increased by controlling the dipole-dipole interaction, determines the time for achieving the maximally entangled state and, the faster the maximally entangled state is prepared, the smaller the environment perturbation.

To illustrate the role of the preparation time in the entanglement dynamics, we plot the maximum value of entanglement between the qubits as a function of $R=J_{12}/J^{exp}_{12}$ for a different preparation time $t_{prep}$ in Fig.~(\ref{fig4}). As expected, such results show an enhancement of the entanglement when $J_{12}$ is increased and when $t_{prep}$ is decreased. This behavior is related to the fact that a maximally entangled configuration is faster achieved for larger $J_{12}$ but the interaction with the environment disturbs the ideal initialization during the preparation time $t_{prep}$. To understand how the preparation time affects the maximally entangled state, in Fig.~ (\ref{fig6}), we plot the maximum value of concurrence as a function of $t_{prep}$ for a fixed $R=50$. One can notice that the maximum concurrence linearly decreases as a function of $t_{prep}$. Such results show what are the parameters to enhance the entanglement in $S$-$T_0$ qubits. 

\section{Conclusion}
In summary, we proposed a model to describe a open quantum system composed of two DQDs. By employing such a model, we were able to compare the case where both DQDs are coupled to a common environment to a situation where each DQD is coupled its own bath of harmonic oscillators. By performing a systematic analysis of  the environment description, we found that the independent environment case agrees with the experimental data shown in ref.~[\onlinecite{sci}]. Moreover, we study the role of the temperature and of the preparation time in the dissipative dynamics. We show that the temperature presents a crucial role in the entanglement evolution. Furthermore, we show that the maximally entangled state decays linearly with the preparation time.

\section{Acknowledgements}
We thank M. Shulman and F. Brito for helpful discussions.
LKC and FFF are grateful to the Brazilian Agencies FAPESP (grants 12/13052-6 and 12/50464-0), CNPq (grants 308839/2012-9 and 474592/2013-8), and CAPES for financial support. KB and FFF would like to thank the International Centre for Theoretical Physics (ICTP) for financial support.

\end{document}